# Molecular circuits based on graphene nano-ribbon junctions

Zhiping Xu[†]

*Department of Engineering Mechanics, Tsinghua University, Beijing, 100084, China*

Graphene nano-ribbons junctions based electronic devices are proposed in this Letter. Non-equilibrium Green's function calculations show that nano-ribbon junctions tailored from single layer graphene with different edge shape and width can act as metal/semiconductor junctions and quantum dots can be implemented. In virtue of the possibilities of patterning monolayer graphene down to atomic precision, these structures, quite different from the previously reported two-dimensional bulk graphene or carbon nanotube devices, are expected to be used as the building blocks of the future nano-electronics.

---

[†] Email: xuzhiping@gmail.com

Nano-electronics, or molecular electronics have been proposed as the alternative to silicon in future technical applications[1] and have attracted great interests recently. In virtue of their unique structures and various functions, these nanostructures possess intriguing electromagnetic, mechanical and optical features. Especially, carbon based nanostructure, such as fullerene, graphene and carbon nanotubes are the most interesting structures because of their rich variety of excellent physical properties. For instance, anomalous quantum hall effects (QHE) and massless Dirac electronic behavior have been discovered in the graphene systems[2,3], and these discovery has sparked lots of investigations on this unique two-dimensional material. Tailored from monolayer graphene, graphene ribbon (GNR) with finite width has been shown to hold unusual electronic properties[4], depending on their edge shape and width. In more details, ribbons with zigzag edges (ZGNRs) possess spin-polarized peculiar edges states and spin-polarized electronic state provides half-metallicity under transverse electric field and has great potential in the application as spintronics[5]. In contrast, the armchair edged ribbons (AGNR) can be either metallic or semiconducting depending on their width[6], AGNR with width $N_a$ (named as $N_a$AGNR in the conventional nomenclature) has been shown to be metallic only if $N_a = 3k + 2$ and semiconducting otherwise, where $k$ is an integer.

From the experimental point of view, the fascinating feature of the ribbons is that the graphene material can be easily patterned using standard micro- or nano-electronics lithography methods. Unlike the carbon nanotubes or other low-dimensional nanostructures, the GNRs with intricate sub-micrometer structures can now be fabricated[7,8,9], and it is believed that a combination of standard lithographic and chemical methods will help to pattern the graphene with atomic precision down to the molecular level. The high mobility $\mu = 2.7$ m$^2$/V·s, large elastic mean free path $l_e = 600$nm and phase coherence lengths $l_\varphi = 1.1$ μm observed[7] in the epitaxial graphene patterned suggest the use of pure GNR structures as the building blocks of the nanoscale confined and coherent electronic circuits. To realize the components such as field transistors[9] and coulomb blockade devices, experimentally controllable metal/semiconductor junctions and quantum dots will be essential. As proposed by Chico *et al.*[10,11], these can be achieved by jointing different carbon nanotubes. However, the fabrication and control of the nanostructure of graphene ribbons are much more convenient than introducing pentagon-heptagon defects in carbon nanotubes as discussed, therefore it is interesting to investigate the possibilities of ribbon junction based nano-circuits.

To this end, we have proposed several kinds of the GNRs based electronic devices in this Letter. We show that, by controlling the tailoring process of GNRs with different edge shape and width, the metal/semiconductor junctions and quantum dots can be easily implemented experimentally. To validate this, electronic transport calculation using the non-equilibrium Green's function method have been carried out following Landauer's approach[12]. The electronic structure of the graphene lattice is described using the nearest-neighbour $\pi$-orbital tight-binding model and the hopping parameter $V_{pp\pi}$ = 2.75 eV is used. This simple topological model gives quantitative results comparing with the LDA results except for the gap opening at small width as the consequence of the length changing of $\sigma$ bonds[6]. By solving the Green's function, the conductance was finally calculated as $G = G_0\text{Tr}[\mathbf{\Gamma}_L\mathbf{G}^R\mathbf{\Gamma}_R\mathbf{G}^A]$ and the density of state is expressed as $D = -\text{Im}\text{Tr}[\mathbf{G}^R]/\pi$[11], where $G_0 = 2e^2/h$ is the unit quanta of conductance including the spin degeneracy, $\mathbf{G}^{R(A)}$ is the retarded (advanced) Green's function of the conductor and $\mathbf{\Gamma}_{L(R)}$ is the spectral density describing the coupling between the left (right) lead and the conductor. In our model, the leads are represented using semi-infinite graphene ribbons attached to the conductor region, with the same shape and width.

First of all we investigate straight metal/semiconducting junction 11AGNR/10AGNR. The structure of the junction is considered by simply patching two different straight ribbons together, leave a width mismatching at the interface. The result shown in Fig. 1 indicates a gap $E_g$ = 0.93 eV near the Fermi energy and the imperfection at the interface induces a deviation of conductance from the step-like curve of the perfect ribbon. However, the van Hove singularities which are the characteristics of 1D system remain.

To examine the detailed electronic structure of the junction, a spatial-resolved localized density of states (LDOS) analysis is helpful. We have grouped the atoms into slices according to their distance from the interface. Each 4.26 Å long slice (a unit cell of the perfect AGNR) in the 10AGNR, 11AGNR and interface part contain 20, 22 and 21 atoms respectively. The LDOS averaged at different slices are plotted in Fig. 1. From the semiconducting 10AGNR side we find the LDOS is distorted near the interface and gap state appears through the contact with metallic 11AGNR. However at slices far from the interface, at slice 3 for example, the perfect semiconduting behavior is mostly recovered. At the scattering interface the van Hove singularities have been smoothed and 1D metallic structure gradually emerges as the distance from interface increases from the 11AGNR side. The arising of gap state near the interface characterizes the metal-semiconductor junction and suggests the possibilities of building Schottky devices.

Furthermore, $L$-shape GNR junctions with different orientations can be constructed. For instance, the LDOS of 8ZGNR/15AGNR junction with a $\pi/6$ joint is analysized in Fig. 2. As expected, the edge state of the 8ZGNR spreads into the semiconducting 15AGNR side. Because of the ZGNR possess spin-polarized structure, so this half-metal/semiconducting junction inspires interests in the spin-transport devices.

Beside of the metal/semiconductor junction, the semiconductor/semiconductor junctions have also been investigated and defect states in gap appear at the interface. Moreover in the ZGNR/ZGNR junctions, zero-conductance dips[13] near Fermi energy have been observed, caused by the complete backward scattering.

The metal-semiconductor junction also suggests quantum dot devices through combing two of them together. We now consider the junction 12AGNR/11AGNR/12AGNR. In this structure a central metallic ribbon is sandwiched by two semiconductor barriers where quantized states can be formed. Our calculation results depicted in Fig. 3 show two sharp DOS peaks inside the gap of semiconducting 12AGNR containing 7 unit cells, with energy $E_{1,2}$ = 0.2025 and -0.2025 eV. As seen from the spatial-resolved LDOS at $E$ = 0.2025, the bounded state is localized inside the 11AGNR region. The structure of the quantum levels can be further tuned by changing the length of 11AGNR. From our calculation, as it changes from 1 to 8 unit cells, the energy spacing between the nearest peaks around Fermi energy, i.e. $\Delta E = E_1-E_2$, gradually decreases from 0.785 eV to 0.385 eV and their DOS becomes higher and sharper.

We have also observed quantized edge states within the 10AGNR/7ZGNR/10AGNR junctions through introducing two π/6 joints. The results are shown in Fig. 3 where we can found 7 LDOS peaks inside the zero-conductance gap. The quantized states with $E$ = -0.3525, -0.1625, -0.05, 0, 0.05, 0.1625 and 0.3525 correspond to different LDOS patterns (see Fig. 4 for $E$ = 0.3525). The higher the energy, the more nodes of the bounded standing wave have. The electron wave quantized pattern depends on the structure of the central region.

In conclusion, we have proposed nano-electronic circuits based on graphene nano-ribbon junctions. Through tailoring GNRs into junctions of different edge shape and width, we can implement metal/semiconductor junctions and quantum dots in principle. In virtue of the possibility of molecular level patterning based on lithography and chemical methods, these devices are expected to be fabricated easier in comparison with other structures such as the single molecule or carbon nanotubes junctions, and are expected to find great applications in the large-scale integrated nano-circuits in future.

The work is supported by the National Science Foundation of China through Grants 10172051, 10252001, and 10332020 and the Hong Kong Research Grant Council (NSFC/RGC N HKU 764/05 and HKU 7012/04P). ZX also thanks Prof. Wenhui Duan, Dr. Tao Zhou and Dr. Haiyun Qian from the Department of Physics in Tsinghua University for their help on the calculation.

FIG. 1. The metal/semiconducting junction 11AGNR/10AGNR: (Top) Conductance and DOS of the whole system. (Bottom) LDOS at slices near the interface. Slice $n$ ($n$ = 1, 2 and 3) represents the $n$-th nearest slice to the interface and the vertical scale of DOS is 0.2.

FIG. 2. Spatial-resolved LDOS in metal/semiconducting junction 8ZGNR/15AGNR, the vertical scale of DOS is 0.2.

FIG. 3. Quantum dot structure based on 12AGNR/11AGNR/12AGNR junction: (Top) Conductance and DOS at low bias, where two isolated sharp peaks appear inside the gap; (Bottom) Spatial-resolved LDOS at $E$ = 0.2025 eV, the grey dot represents the ionic site and the radius of circle around it corresponds to the value of LDOS.

FIG. 4. Quantum dot structure based on 10AGNR/7ZGNR/10AGNR junction: (Top) Conductance and DOS; (Bottom) Spatial-resolved LDOS at $E$ = 0.3525 eV.

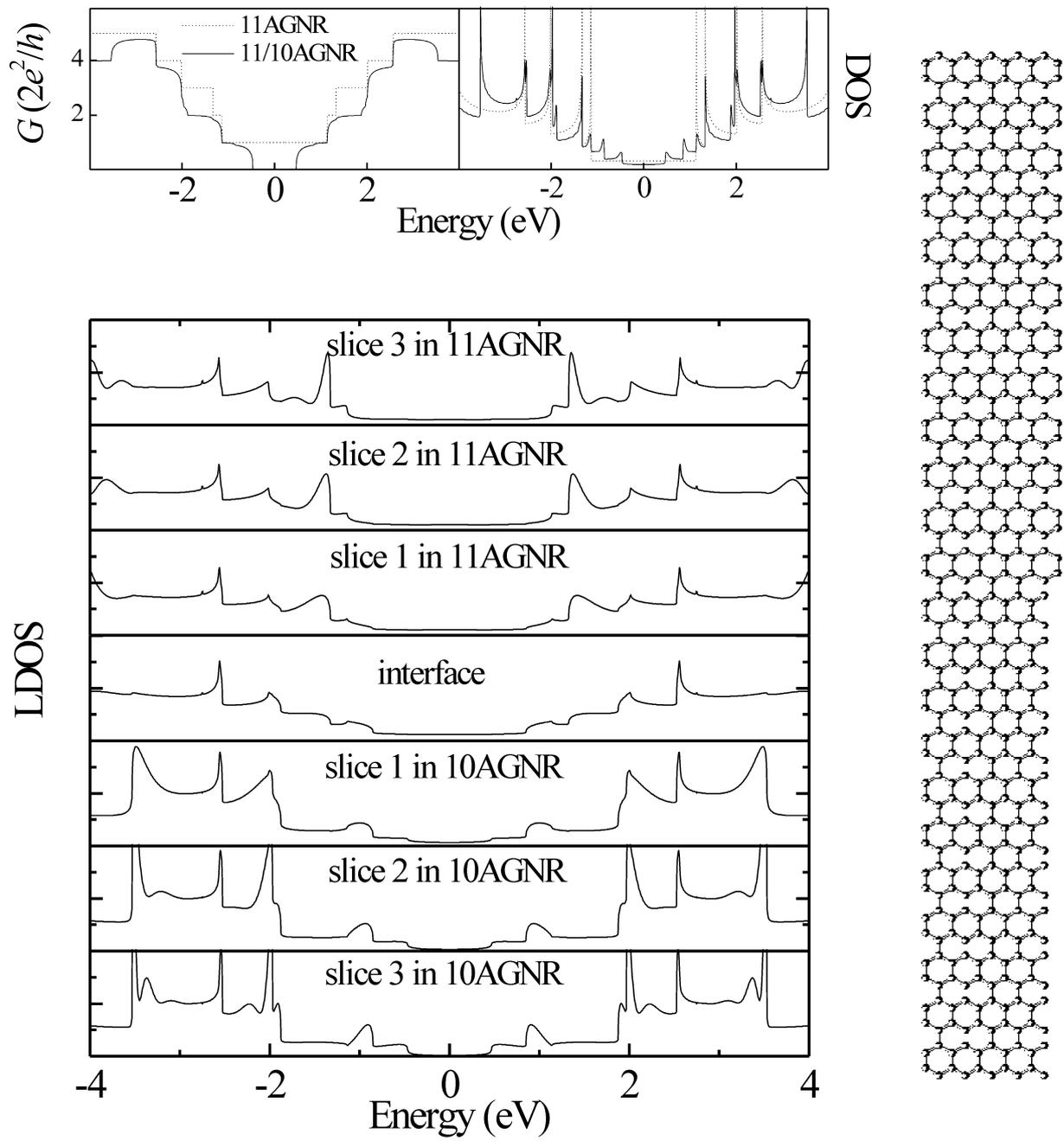

Figure 1

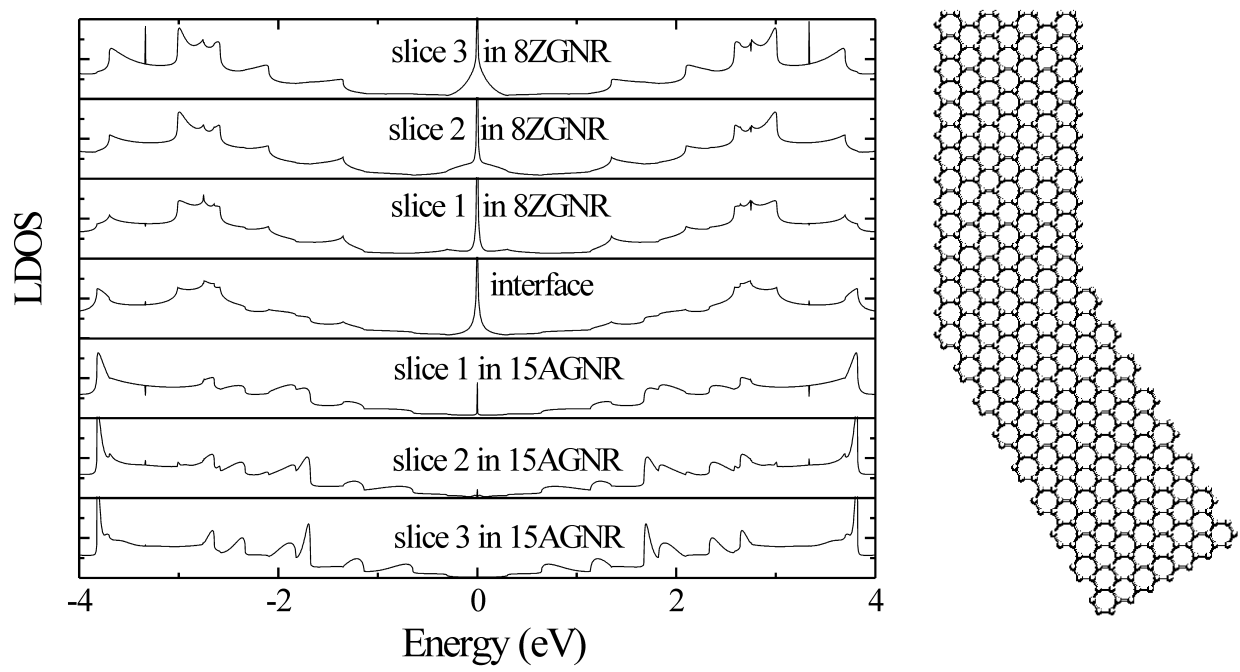

Figure 2

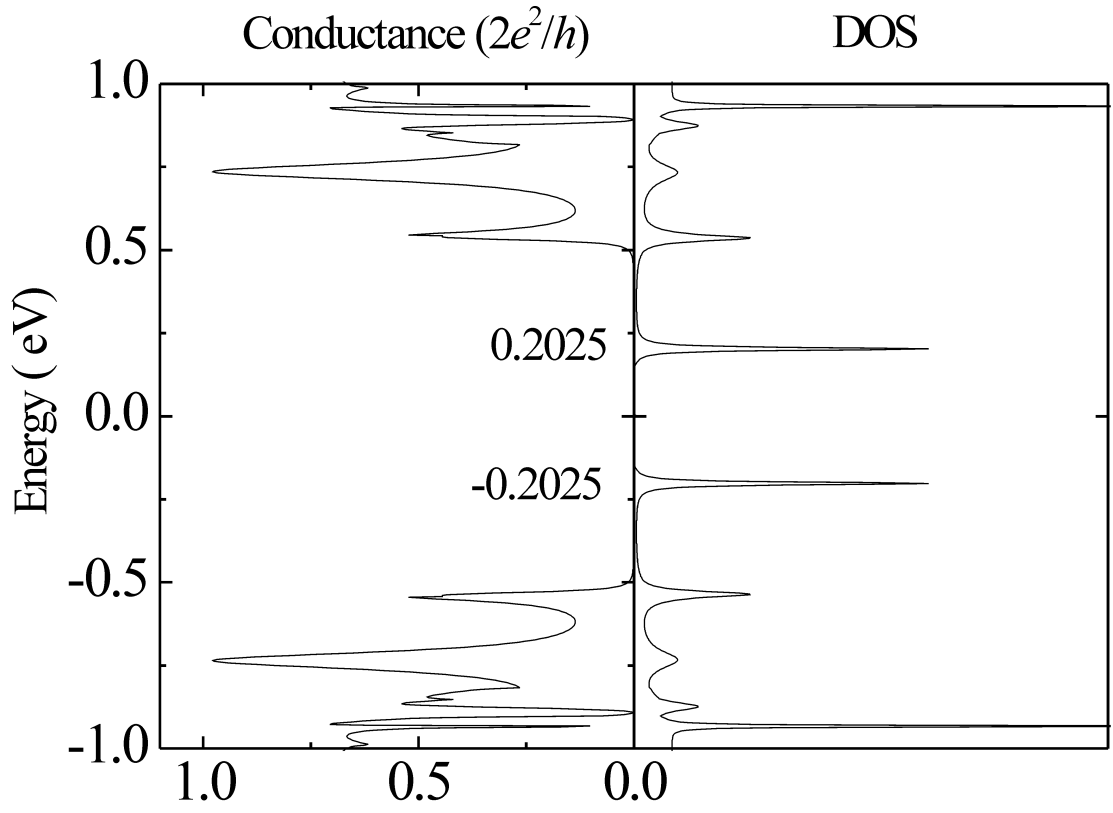

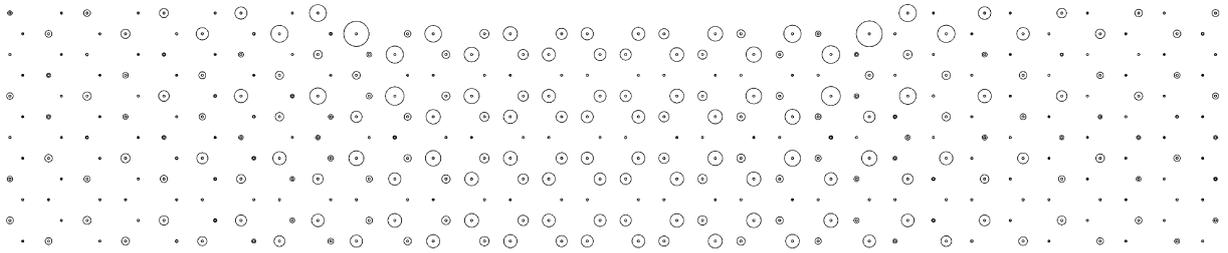

Figure 3

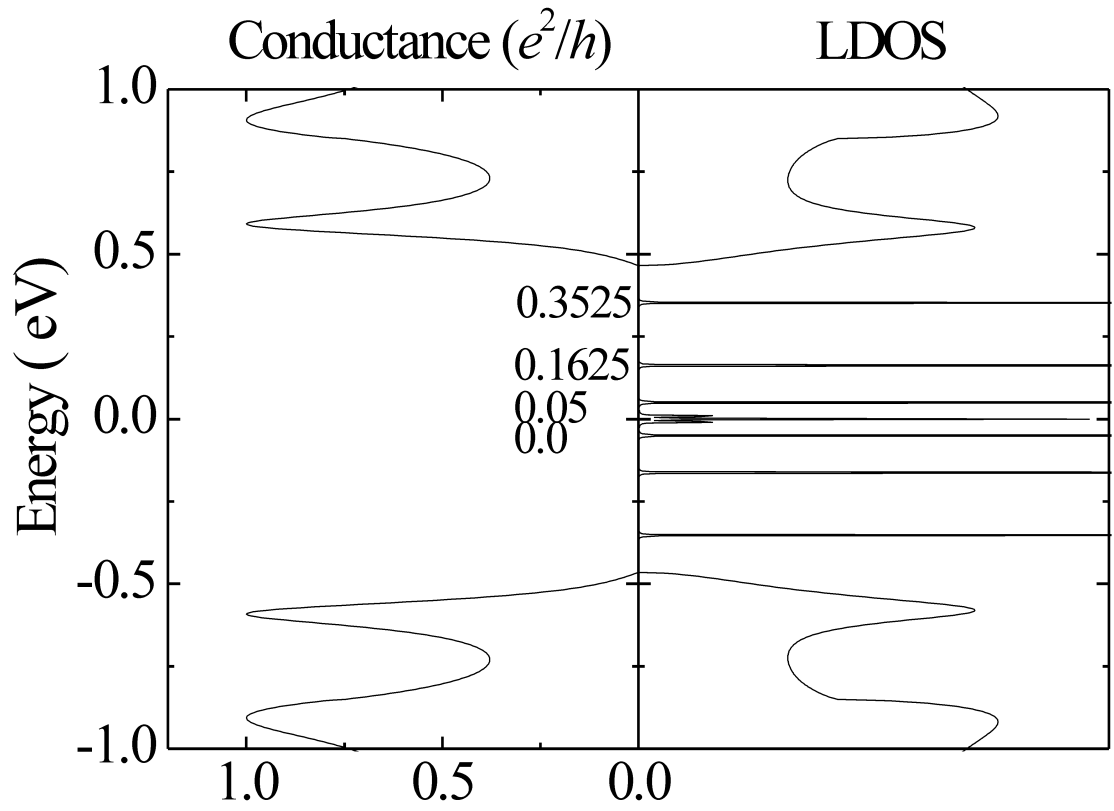

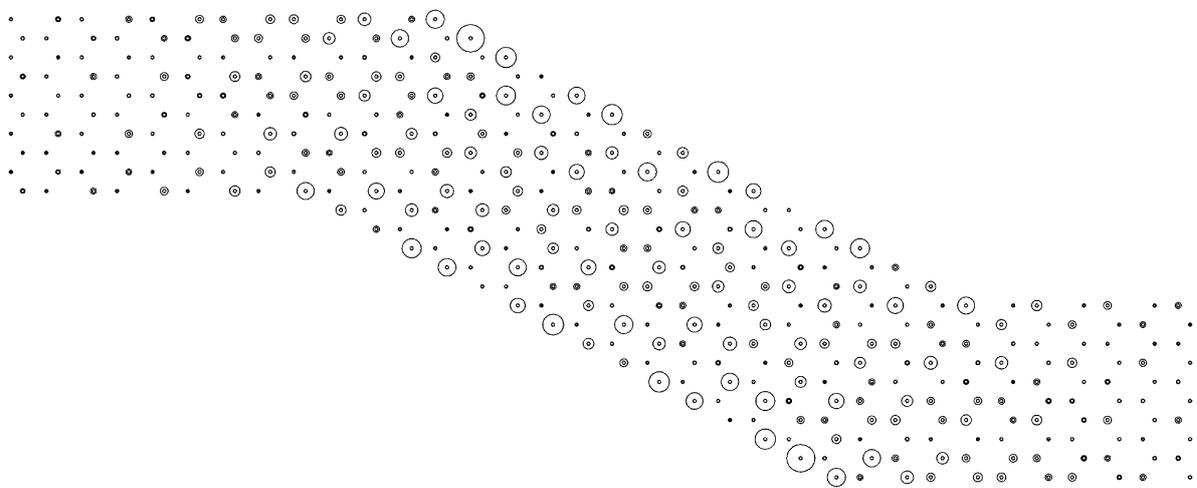

Figure 4